\documentclass[twocolumn,pr,nofootinbib,notitlepage,longbibliography]{revtex4-1}
\pdfoutput=1
\usepackage{amsmath}
\usepackage{amssymb}
\usepackage{bm}
\usepackage{epsfig}
\usepackage{graphicx}
\usepackage{color}
\usepackage[colorlinks]{hyperref}
\usepackage[english]{babel}

\renewcommand{\Re}{\mathop{\rm Re}}

\begin{document}

\title{Intervalley Polaron in Atomically Thin Transition Metal Dichalcogenides}
\author{M.M. Glazov, M.A. Semina} 
\affiliation{Ioffe Institute, 194021 St. Petersburg, Russia}

\author{C. Robert, B. Urbaszek, T. Amand, X. Marie}
\affiliation{Universit\'e de Toulouse, INSA-CNRS-UPS, LPCNO, 135 Av. Rangueil, 31077 Toulouse, France}

\begin{abstract}
We study theoretically intervalley coupling in transition-metal dichalcogenide monolayers due to electron interaction with short-wavelength phonons. We demonstrate that this intervalley polaron coupling results in (i) a renormalization of the conduction band spin splitting and (ii) an increase of the electron effective masses. We also calculate the renormalization of the cyclotron energy and the Landau level splitting in the presence of an external magnetic field. An inter-valley magneto-phonon resonance is uncovered. Similar, but much weaker effects are also expected for the valence band holes. These results might help to resolve the discrepancy between \emph{ab initio} values of the electron effective masses and the ones deduced from magneto-transport measurements.
\end{abstract}

\maketitle

\textit{Introduction.} Fascinating electronic and optical properties of two-dimensional (2D) materials like graphene, transition-metal dichalcogenide (TMD) monolayers (MLs), atomically-thin hexagonal boron nitride, black phosphorous, and others have attracted strong interest~\cite{Geim2007,Geim:2013aa,ajayan2016van}. Unusual band structure of TMD MLs with two valleys in conduction and valence bands, where the spin degeneracy of the electron and hole states is removed and the spin and valley degrees of freedom are locked~\cite{Xiao:2012cr,Xu:2014cr} has made  2D TMD crystals extremely attractive for study of spintronic and valleytronic effects~\cite{Song:2013uq,dery2015polarization, Wang:2017b,dey2017gate}. Chiral selection rules combined with strong excitonic effects provide unprecedented access to spin and valley indices of charge carriers by optical means~\cite{RevModPhys.90.021001,Durnev_2018}. For in-depth studies of TMD MLs and the development of possible applications,  the basic band structure parameters, including band gap and effective masses should be reliably established both from experiments and modeling.

Optical spectroscopy has made it possible to determine key parameters such as exciton band gap,  binding energy,  Land\'e factor and evaluate the reduced masses of the electron-hole pairs from the high-magnetic field experiments~\cite{RevModPhys.90.021001,Stier:2018a}.
The single electron parameters, namely, the effective masses of charge carriers and their individual $g$ factors are usually hard to determine optically. These quantities are, as a rule, inferred from transport measurements. Thanks to the improvement of carrier mobility obtained in hBN encapsulated TMD monolayers~\cite{cadiz2017excitonic,Gustafsson:2018aa}, quantum transport measurements have been recently performed. 

Observation of Landau levels has allowed one to determine the effective masses of the charge carriers in the valence and conduction band~\cite{PhysRevLett.116.086601,PhysRevB.97.201407,Gustafsson:2018aa,PhysRevLett.121.247701,Lin:2019aa}. Interestingly, while the measured valence band hole (VB) masses are in good agreement with both first-principles modeling and angle resolved photoemission spectroscopy (ARPES) measurements~\cite{Zhang:2013aa,jin2013direct}, the discrepancies for the conduction band (CB) effective masses in Mo-based monolayers are substantial~\cite{PhysRevLett.121.247701,PhysRevB.97.201407} and can hardly be related to electron-electron interaction and inaccuracies in density functional theory (DFT) calculations. The transport measurements in both MoS$_2$ and MoSe$_2$ MLs yield characteristic CB electron effective mass of $0.7$---$0.8m_0$, typically twice larger than the ones deduced from DFT calculations ($m_0$ is the free electron mass)~\cite{2053-1583-2-2-022001,wickramaratne2014electronic}.

\begin{figure}
\includegraphics[width=0.9\linewidth]{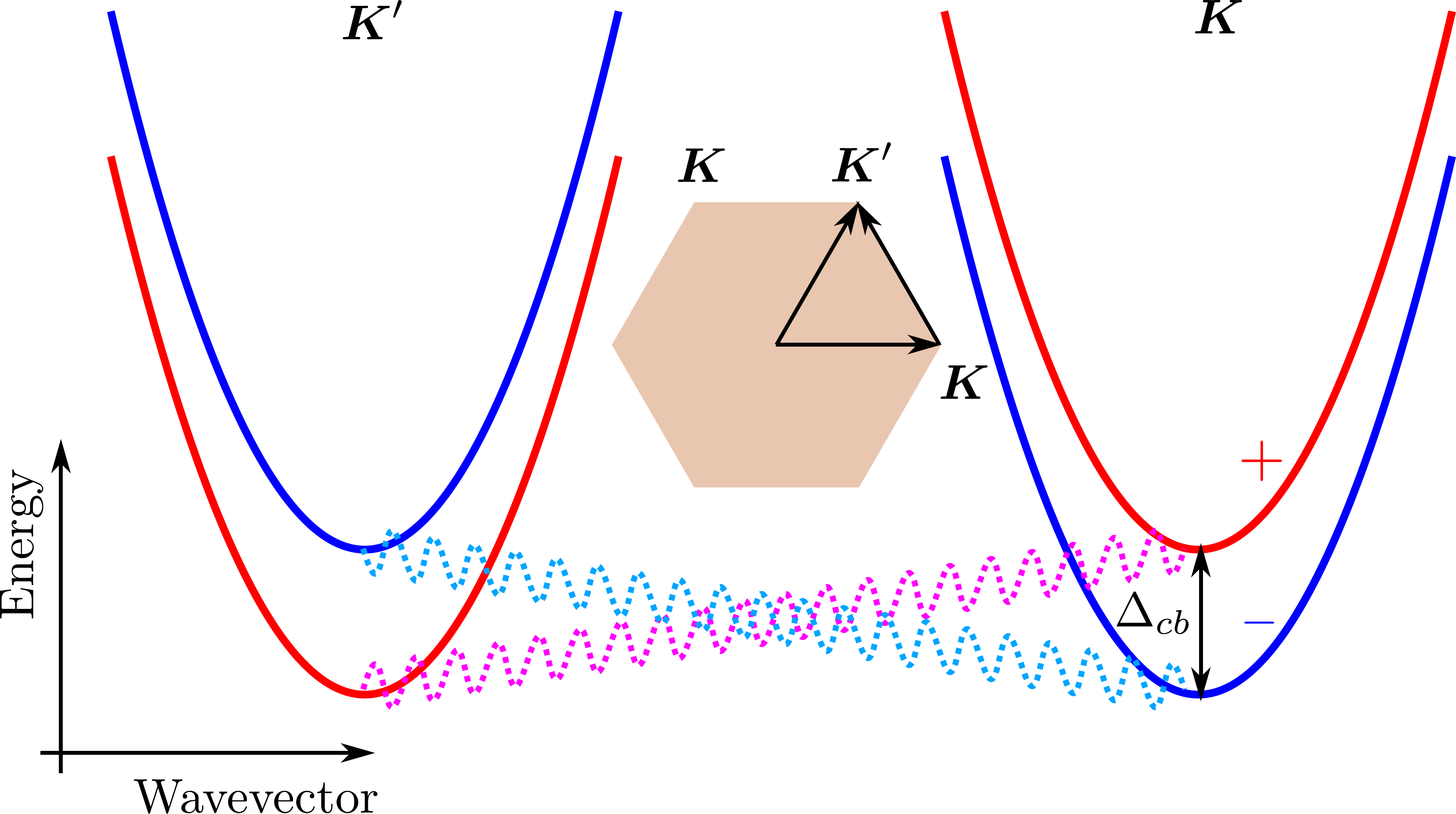}
\caption{Schematic illustration of the conduction subbands in the vicinity of the $\bm K$ and $\bm K'$ points at the BZ edge. Inset shows hexagonal BZ. Wiggly lines illustrate phonon-induced coupling between the valleys.   Signs ``$+$'' and ``$-$'' denote top and bottom subbands.}\label{fig:scheme}
\end{figure}

Here we demonstrate a pathway which could improve the agreement between the theory and the experiment for conduction band masses in TMD MLs. We show that two valleys $\bm K$ and $\bm K'$ can be coupled by the phonons with  wavevectors close to the edge of the Brillouin zone (BZ). Let us recall that the spin up and the spin down CB bands in each non-equivalent valley $\bm K$ and $\bm K'$ are split by the spin-orbit (SO) interaction, with a typical energy difference, $\Delta_{cb}$, of a few tens of meV, see Fig.~\ref{fig:scheme}~\cite{kosmider2013large,2053-1583-2-2-022001}. The intervalley electron-phonon coupling is stronger for the CB because electron-phonon interaction is largely spin-conserving, thus strong SO splitting of VB (of the order of hundreds of meV) suppresses the mixing. We develop an analytical model for this new type of polaron and demonstrate that the intervalley polaron effect results in the renormalization of both the CB spin splitting and the effective masses of the electrons. We discuss also the renormalization of the Landau level energies due to the intervalley electron-phonon interaction. Our simple model indicates an increase of the electron effective mass due to inter-valley polaron, in qualitative agreement with transport measurements, but the quantitative agreement with the experiments requires exaggerated coupling constant possibly due to oversimplification of the model.

\textit{Intervalley electron-phonon interaction.} Electrons propagating through the crystal interact with the host lattice atoms and produce  local deformations of the lattice. Thus, the periodic potential experienced by the electron is modified resulting in a modification of the electron energy, and the dragging of the crystal lattice deformation with the electron gives rise to an increase of electrons effective mass as a result of the polaron effect~\cite{landau:polaron,pekar1946local,landau:pekar,doi:10.1080/00018735400101213,Alexandrov2010}. This effect was found to be important to explain effective mass variations in many different semiconductors~\cite{PhysRevB.54.1467,PhysRevB.57.3966,PhysRevLett.74.1645,PhysRevLett.100.226403}. Quantum mechanically, electron-phonon interaction due to virtual processes of emission and reabsorption of the phonon by the electron gives rise to the formation of the phonon cloud surrounding the electron and renormalizes its dispersion. In the second-order perturbation theory the electron self-energy, $\Sigma_{\bm p}(\varepsilon)$, due to the coupling with phonons can be written at zero temperature as
\begin{equation}
\label{self}
\Sigma_{\bm p}(\varepsilon) = \sum_{\bm q,\alpha;j} \frac{|M_{\bm q,\alpha}^j|^2}{\varepsilon - \hbar\Omega_{\bm q,\alpha}  - E_{\bm p - \bm q}^{j} + \mathrm i \delta}.
\end{equation}
Here $\bm p$ is the electron wavevector in the crystal, $\varepsilon$ is the energy variable, $\bm q$ is the phonon wavevector, $\alpha$ enumerates the phonon branches, $j$ enumerates the intermediate states of the electron,  $M_{\bm q,\alpha}^j$ is the electron-phonon coupling matrix element, $\Omega_{\bm q,\alpha}$ is the frequency of the corresponding phonon mode and $E_{\bm p}^j$ is the electron dispersion in the state $j$, the term with $\delta\to +0$ in the denominator ensures the causality. The self-energy~\eqref{self} describes the electron energy and dispersion renormalization in the second order in the electron phonon interaction.

Leading contributions to Eq.~\eqref{self} are provided mainly by nearest intermediate states  and also by high-symmetry points of the BZ where the phonon dispersion is flat and the density of states is increased~\cite{christiansen2017phonon}. Analysis of the TMD MLs CB structure shows that the main contributions to the self-energy are produced by the intermediate states in the same band and the same valley, i.e., by the standard, \emph{intravalley}, Fr\"ohlich-like polaron, studied, e.g., in Refs.~\cite{doi:10.1063/1.5030678,doi:10.1063/1.5025907,2019arXiv190102567V} as well as by the intermediate states in the opposite valley ($\bm K'$ for the electrons in the $\bm K$ valley and vice versa), which result in the \emph{intervalley} polaron, not investigated so far. Key difference for the intervalley polaron compared with the intravalley polaron is the presence of the CB spin splitting in the energy denominator. 

The results on intravalley polaron are summarized in the supplementary information (SI), which also contains justification of applicability of the second-order perturbation theory to calculate $\Sigma_{\bm p}(\varepsilon)$ in Eq.~\eqref{self} for the TMD ML system~\cite{supp}. Here we focus on the new intervalley polaron which results in the coupling of electron states in $\bm K$ and $\bm K'$ valleys as schematically shown in Fig.~\ref{fig:scheme}. Generally, in order to calculate the self energy, the electron and phonon dispersions across the whole BZ are needed; here we take into account the contribution coming from the vicinity of $\bm K'$ (for $\bm K$ valley electron) where the density of states is largest. We focus on the renormalization of electron states in the $\bm K$ valley, correspondingly, we replace $M_{\bm q,\alpha}^i$ by $M_{\bm K,\alpha}^i$, its value for $\bm q = \bm K$, we also disregard the phonon dispersion replacing $\hbar\Omega_{\bm q,\alpha}$ by $\hbar\Omega_{\bm K,\alpha}$. In what follows we omit the subscript $\alpha$ (keeping in mind that in the final result one has to sum over all appropriate phonon modes) and take into account that the spin is conserved by the electron-phonon interaction. Changing the integration variable from $\bm q$ to $\bm q' = \bm q - \bm K$,  and taking into account that for the electron in the top (bottom) spin subband the intermediate state energy reads $E^{j}_{\bm p - \bm  q} \approx  \mp \Delta_{cb} + E_{\bm p - \bm q'}$, $E_{\bm p} = \hbar^2p^2/(2m)$, where $m$ being the bare CB effective mass~\cite{mass}, 
we arrive at 
\begin{multline}
\label{self:inter:1}
\Sigma_{\bm p, \pm}(\varepsilon) = -\frac{\beta_{\bm K} (\Delta_{cb} + \hbar\Omega_{\bm K})}{4\pi} \times\\ 
 \int_0^{E_Q} \frac{dE }{\sqrt{(\varepsilon - E_{\bm p} \pm \Delta_{cb} - \hbar\Omega_{\bm K} - E)^2-4E_{\bm p} E}}.
\end{multline}
Here subscripts correspond to the top ($+$) and bottom ($-$) spin subbands, $\Delta_{cb}>0$. An effective dimensionless intervalley coupling constant in Eq.~\eqref{self:inter:1} reads
\begin{equation}
\label{beta:inter}
\beta_{\bm K} = \frac{2\mathcal S m|M_{\bm K}|^2}{\hbar^2(\Delta_{cb} + \hbar\Omega_{\bm K})},
\end{equation}
and the cut-off energy  $E_Q = \hbar^2 Q^2/2m$ with the cut-off wavevector $Q\sim |\bm K|$ is needed to avoid the logarithmic divergence of the integral, $\mathcal S$ is the normalization area. We have omitted infinitesimal term $\mathrm i \delta$ in the denominator for brevity. It is worth stressing that $\beta_{\bm K}$ depends not only on the matrix element of the electron-phonon intervalley interaction but also on the combination of the CB spin splitting and the phonon energy $\Delta_{cb}+\hbar\Omega_{\bm K}$. The logarithmic divergence does not mean particularly strong electron-phonon coupling,  it demonstrates the limitations of analytical approach with parabolic bands, flat dispersion of phonons and momentum-independent coupling. 

\begin{figure}
\includegraphics[width=0.9\linewidth]{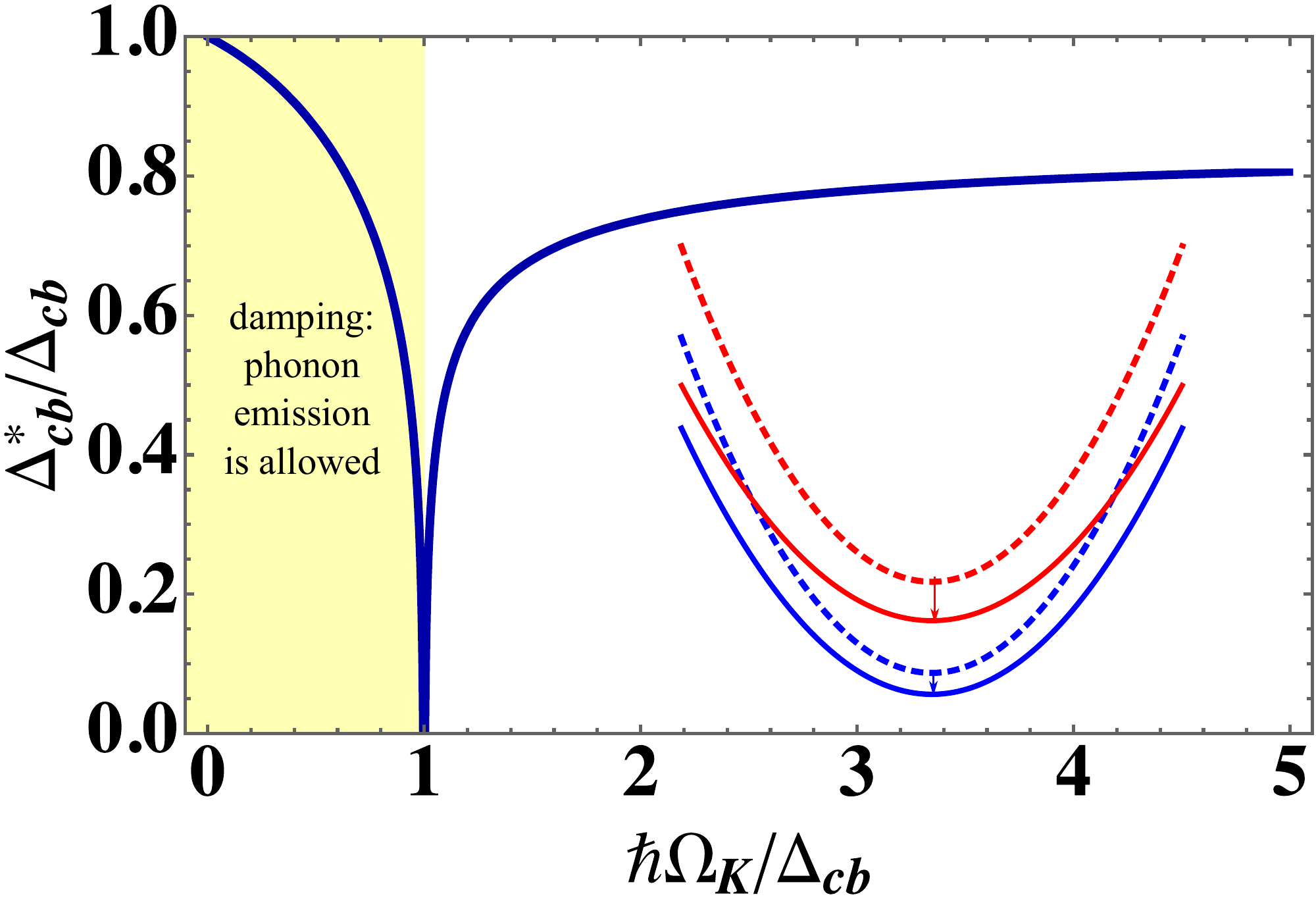}
\caption{Renormalized conduction band spin splitting as a function of the ratio between the phonon energy $\hbar\Omega_{\bm K}$ and bare conduction band spin splitting $\Delta_{cb}$ calculated for $\beta_{\bm K}=1$ after Eq.~\eqref{renorm:delta:cb:1}. Filled area corresponds to the situation where $\hbar\Omega_{\bm K}< \Delta_{cb}$ and the real phonon emission processes are possible. Inset demonstrates relative shifts of upper and lower spin subbands in one valley due to the electron-phonon interaction with another valley and the renormalization of the effective mass (exaggerated for illustration).}\label{fig:Delta}
\end{figure}

\textit{Renormalization of band parameters.} The electron-phonon interaction produces noticeable effect on conduction subband energies~\cite{shift}. 
We predict a significant renormalization effect of the CB spin splitting. Indeed, as it is seen from Fig.~\ref{fig:scheme}, for the electron in the lower spin subband the intermediate state has a higher energy, $\Delta_{cb}+\hbar\Omega_{\bm K}$, while for the electron in the higher spin subband the intermediate state has a lower energy, $-\Delta_{cb}+\hbar\Omega_{\bm K}$. Under the condition $\hbar\Omega_{\bm K} > \Delta_{cb}$ both subbands are pushed by the electron-phonon interaction towards the lower energies, but the shift of the upper subband is larger, see inset in Fig.~\ref{fig:Delta}. 
The variation of the CB spin splitting can be recast as
\begin{equation}
\label{renorm:delta:cb}
\delta \Delta_{cb} = \lim_{p\to 0}\left[\Sigma_{\bm p,+}(E_{\bm p}) - \Sigma_{\bm p,-}(E_{\bm p})\right].
\end{equation}
We introduce the renormalized CB spin splitting $\Delta_{cb}^* = \Delta_{cb} + \Re\{\delta \Delta_{cb}\}$ and present it in the form
\begin{equation}
\label{renorm:delta:cb:1}
\Delta_{cb}^*=\Delta_{cb} \left[1-\frac{\beta_{\bm K}}{4\pi} \left(1+ \frac{\hbar\Omega_{\bm K}}{\Delta_{cb}}\right) \ln{\left(\frac{\hbar\Omega_{\bm K}+\Delta_{cb}}{ \hbar\Omega_{\bm K} - \Delta_{cb}}\right)} \right].
\end{equation}
Note that logarithmic divergencies in the self-energies cancel and the result~\eqref{renorm:delta:cb:1} does not depend on the cut-off energy.
For $\Delta_{cb} > \hbar\Omega_{\bm K}$ the argument of the logarithm becomes negative and, formally, $\Delta_{cb}^*$ acquires an imagninary part. In this case, the energy shift is given by the real part of the same expression~\eqref{renorm:delta:cb:1}. However, the intervalley transitions from the upper to the lower subband accompanied by the phonon emission become possible. The associated damping of the electron in the upper spin subband is determined by the imaginary part of $\Delta_{cv}^*$. The calculated renormalization of the CB spin splitting is plotted in Fig.~\ref{fig:Delta} as a function of the ratio $\hbar\Omega_{\bm K}/\Delta_{cb}$. The singularity in the $\Delta_{cb}^*$ at $\hbar\Omega_{\bm K} = \Delta_{cb}$, i.e., at the threshold of the phonon emission, results from the resonance condition where the energy of the electron in the top subband minus phonon energy equals the energy of the electron in the bottom subband. The detailed analysis of this singular behavior can be carried out by the methods developed in Ref.~\cite{Levinson1973}; it is beyond the scope of the present work.

Similarly, one can extract the $p^2$ contribution in the self-energy and obtain the renormalization of the effective mass resulting from the intervalley polaron. We obtain for the top ($+$) and bottom ($-$) spin subbands
\begin{equation}
\label{mass:inter}
m^*_\pm = m\left(1+ \frac{\beta_{\bm K}}{4\pi} \frac{\hbar\Omega_{\bm K}+ \Delta_{cb}}{\hbar\Omega_{\bm K}\mp \Delta_{cb}} \right).
\end{equation}
Interestingly, the polaron effect does not only increase the effective masses but  also leads to a difference of the mass of the two spin subbands:  $m_+^*\ne m_-^*$. Similarly to the renormalization of the $\Delta_{cb}$ the intervalley polaron effect on the masses is singular at $\hbar\Omega_{\bm K} = \Delta_{cb}$, i.e., at the threshold of the phonon emission. Interestingly, the intervalley polaron effect makes both spin subbands heavier and the renormalization of the mass in the topmost subband is larger, $m_+^*> m_-^*$ (inset in Fig.~\ref{fig:Delta}).
Note that the interaction of the CBs with the VBs ($\bm k\cdot\bm p$ mixing) also yields a slight difference difference of masses for top and bottom CBs; this difference is controlled by the dimensionless $\sim \Delta_{vb}/E_g$ where $E_g$ is the band gap and $\Delta_{vb}$ the SO splitting of the VB~\cite{2053-1583-2-2-022001,Durnev_2018}. In contrast to the intervalley polaron effect, this $\bm k\cdot\bm p$-mixing of the bands together with the SO coupling makes topmost subband heavier in Mo-based MLs and lighter in W-based MLs due to different order of spin states~\cite{2053-1583-2-2-022001}. 

\textit{Renormalization of Landau levels.} Let us now discuss the effect of the intervalley polaron on the electron spectrum in magnetic field. We assume that the magnetic field is applied perpendicular to the ML plane and disregard the electron Zeeman splitting assuming it to be much smaller than both the phonon energy $\hbar\Omega_{\bm K}$ and the CB SO splitting $\Delta_{cb}$. As a result, in the absence of electron-phonon interaction, the electron energy spectrum consists of the well-known series of Landau levels 
\begin{equation}
\label{Landau}
E_n = \hbar\omega_c\left(n+\frac{1}{2}\right),
\end{equation}
where $\omega_c = |eB_z/mc|$ is the electron cyclotron frequency.
We calculate the correction to the energy of $n$-th Landau level in the second order in the intervalley electron-phonon interaction. Similarly to Eqs.~\eqref{self} and \eqref{self:inter:1} we obtain (cf. Ref.~\cite{doi:10.1063/1.5025907} where intravalley Fr\"ohlich magnetopolaron in TMD ML was studied)
\begin{equation}
\label{Landau}
\delta E_{n,\pm} = \sum_{\bm q', n'k_y'} \frac{|M_{\bm K}|^2 \left|\langle n'k_y'|e^{\mathrm i \bm q' \bm r}|n k_y\rangle\right|^2}{\hbar\omega_c (n-n') - \hbar\Omega_{\bm K} \pm \Delta_{cb} }.
\end{equation}
Here we disregarded the wavevector dependence of the intervalley matrix element and phonon dispersion, $\pm$ refers to the electron in the top and bottom spin subbands, $\bm q' = \bm q- \bm K$ with $\bm q$ is the phonon wavevector, $|n k_y\rangle$ is the electron state in the Landau gauge, $n$ is the Landau level number, $k_y$ is the component of the electron inplane wavevector. 
We obtain the renormalized electron spectrum in the form of Eq.~\eqref{Landau} with the effective cyclotron frequency $\omega_{c,\pm}^*(n)$ which depends now on the Landau level number and on the spin subband~\cite{supp}:
\begin{equation}
\label{Landau:2}
\frac{\omega_{c,\pm}^*(n)}{\omega_c} = 1 -\frac{\beta_{\bm K}}{4\pi(1-\eta_{\pm} n)}\frac{\hbar\Omega_{\bm K}+ \Delta_{cb}}{\hbar\Omega_{\bm K}\mp \Delta_{cb}},
\end{equation}
where $\eta_\pm = \hbar \omega_c/(\hbar\Omega_{\bm K} \mp \Delta_{cb})$ characterizes the magnetic field strength. 

\begin{figure}
\includegraphics[width=0.9\linewidth]{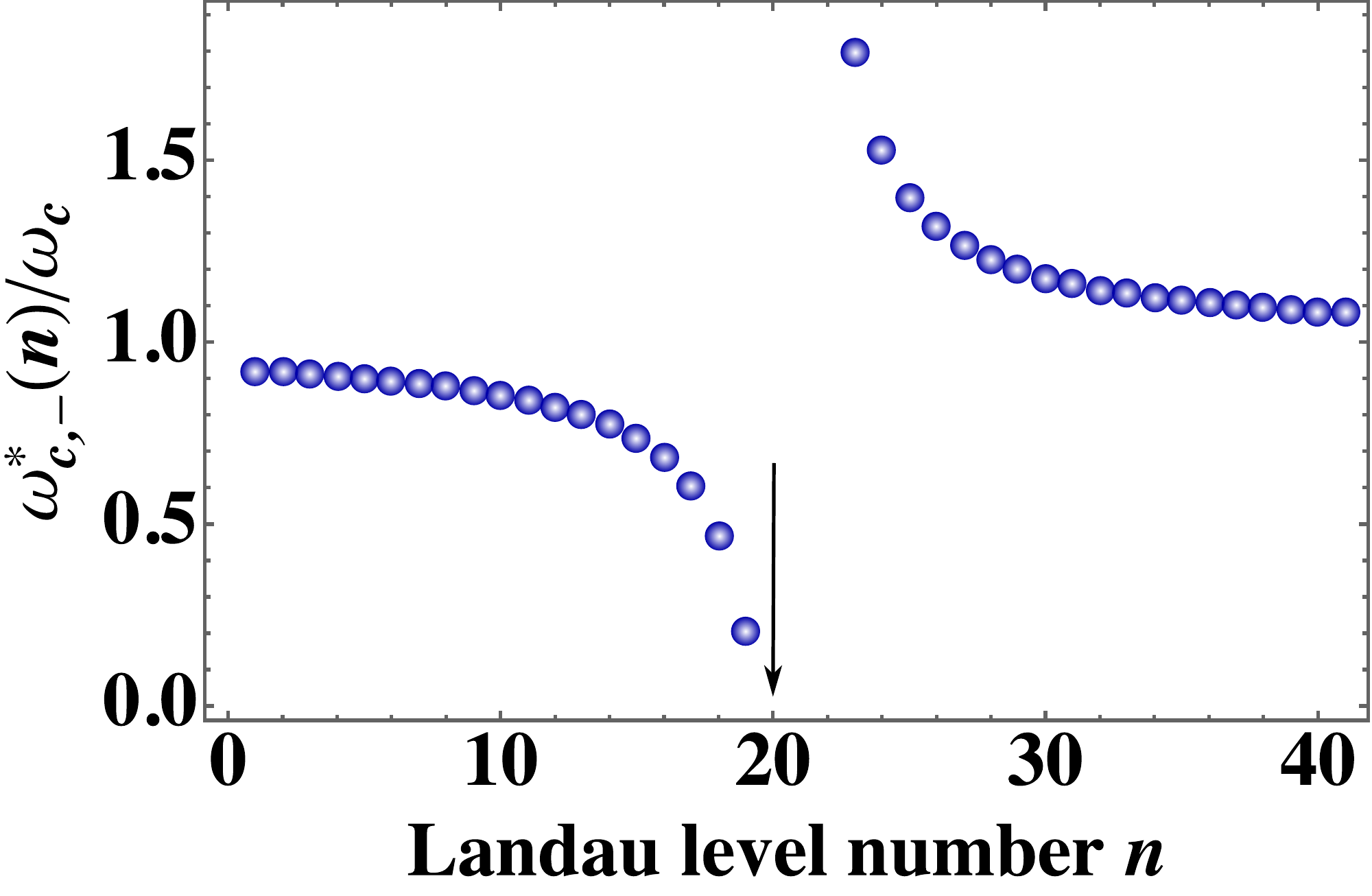}
\caption{Renormalized cyclotron frequency for the bottom spin subband at $\beta_{\bm K}=1$ and $\hbar\omega_c/(\hbar\Omega_{\bm K}+ \Delta_{cb}) = 1/20$ calculated after Eq.~\eqref{Landau:2}. The arrow indicates intervalley magneto-phonon resonance condition, Eq.~\eqref{resonance}.}\label{fig:omegaC}
\end{figure}

In the limit of weak magnetic fields and low Landau levels, where $\eta_\pm n\ll 1$, we have series of equidistant Landau levels in each subband with the renormalized effective mass given by Eq.~\eqref{mass:inter}. For $\eta_\pm n \gg 1$ one recovers unperturbed Landau Levels, Eq.~\eqref{Landau}, because phonon cloud cannot follow the electron orbiting with high enough frequency. Strong renormalization of the cyclotron frequency occurs for the top spin subband at $\hbar\Omega_{\bm K} = \Delta_{cb}$ where resonant emission of phonons becomes possible. Another strong feature in $\omega_{c,\pm}^*(n)$ occurs at 
\begin{equation}
\label{resonance}
n = \left\lfloor \frac{1}{\eta_\pm} \right\rfloor,
\end{equation} 
where $\lfloor \ldots \rfloor$ stands for the integer part. Under this condition there is an intervalley magneto-phonon resonance (cf. Refs.~\cite{gurevich:firsov,PhysRevLett.12.660}): The energy of $n$-th Landau level in one of the subbands is at resonance with the bottom of another subband in the opposite valley. In this case the polaron effect is particularly strong.

\textit{Discussion.} Intervalley polaron coupling in TMD MLs can be enabled by the phonons with large wavevectors $|\bm q| \approx |\bm K|$~\cite{PhysRevB.84.155413,PhysRevB.85.115317,PhysRevB.90.045422,Song:2013uq}. {The symmetry of relevant phonon modes is discussed in SI, for conduction band the main contribution comes from the chiral phonon mode~\cite{Song:2013uq,PhysRevLett.115.115502,Zhu579} whose angular momentum component equals to $\pm 1$ for $\bm K'\rightleftharpoons \bm K$ transfer~\cite{supp}. T}he effect is related to the intervalley deformation potential induced by the lattice vibrations. Note that the Fr\"ohlich interaction is long-range and expected to provide negligible contribution to the intervalley polaron effect. The matrix element $M_{\bm K}$ in Eqs.~\eqref{beta:inter} and \eqref{Landau} can be represented as
\begin{equation}
\label{MK:def}
M_{\bm K} = \sqrt{\frac{\hbar}{2\rho \Omega_{\bm K} \mathcal S}} D_0,
\end{equation}
where $\rho$ is the two-dimensional mass density of the crystal and $D_0$ is the deformation potential parameter, $\mathcal S$ is the normalization area. Substituting Eq.~\eqref{MK:def} into Eq.~\eqref{beta:inter} we obtain a crude estimate for the coupling strength $\beta_{\bm K}$:
\begin{equation}
\label{beta:est}
\beta_{\bm K} \sim \frac{D_0^2}{\hbar\Omega_{\bm K}(\hbar\Omega_{\bm K} + \Delta_{cb})}\frac{m}{\rho} \sim 0.1 \ldots 1,
\end{equation}
at $\rho=3\times 10^{-7}$~g/cm$^2$, $m=0.5m_0$, $D_0 \sim (1\ldots 4)\times 10^8$~eV/cm~\cite{PhysRevB.90.045422} and $\hbar\Omega_{K} \sim \Delta_{cb} \sim 10$~meV. 
With this simplified model and the uncertainty of the parameters involved, we find that the coupling is not particularly strong. It will yield an increase of the bottom CB mass by $\sim 10\%$. We emphasize that the exact values of key parameters such as the CB SO splitting and deformation potential parameters are not very well known. {The CB SO splitting strongly varies for different DFT models~\cite{gerber}.  The deformation potential can be roughly estimated as the ratio of atomic energy scale, $\sim10$~eV, to the lattice constant (several Angstrom) resulting in values consistent with Ref.~\cite{PhysRevB.90.045422}.}
 Additional enhancement is expected for the CB states due to the presence of the $\bm Q$ (also denoted as $\bm \Lambda$) valley. The coupling constant $\beta_{\bm Q}$ of the intervalley polaron involving the $\bm Q$-valley has a similar order of magnitude.

This increase of the CB electron mass in TMD monolayer could qualitatively explain the recent measurements performed on high-quality $n$-type MoS$_2$ and MoSe$_2$ monolayers displaying well resolved Shubnikov de Haas oscillations. These experiments yield CB mass of about $0.7m_0$  in MoS$_2$ MLs and $0.8m_0$ in MoSe$_2$ ML~\cite{PhysRevLett.121.247701,PhysRevB.97.201407}. Surprisingly the masses calculated with DFT approaches are much smaller, typically $(0.4-0.5)m_0$ for both MLs~\cite{2053-1583-2-2-022001,wickramaratne2014electronic}. This can hardly be explained by the electron-electron interaction as the measured electron mass for both systems does not depend much on the doping density. Remarkably this discrepancy between experiment and DFT calculation does not occur for the VB, where both transport~\cite{PhysRevLett.116.086601} and ARPES~\cite{Zhang:2013aa,jin2013direct} measurements yield the effective mass of about $0.4m_0$, in agreement with the DFT calculations~\cite{dens}.
The standard Fr\"ohlich polaron effect could not explain the increase of mass observed for CB only as it should affect in a similar manner both electrons and holes. The same argument rules out possible effects of interfacial polarons due to electron-phonon coupling at the interfaces between the TMD ML and hBN~\cite{chen2018emergence}. 
In contrast, the intervalley polaron effect, as explained above, yields  an increase of the CB mass (compared to the mass calculated with DFT) but produces negligible effect on the VB mass because of much larger VB SO splitting. {Estimates after Eq.~\eqref{beta:est} for the VB yields the value $\beta_{\bm K}^{vb} \sim (\hbar\Omega_{\bm K}+\Delta_{cb})/\Delta_{vb} \beta_{\bm K} \sim \beta_{\bm K}/10$.} Recent magneto-optical spectroscopy of TMD MLs~\cite{crooker:mass:long} reveals an increase in $m$ in Mo-based MLs, indicating the importance of the intervalley polaron effect for magnetoexcitons

Our calculations also predict a significant decrease of the CB SO splitting due to intervalley polaron effect (Fig.~\ref{fig:Delta}). We suggest that the polaron effect should be taken into account when comparing {measured SO~\cite{PhysRevLett.121.247701,PhysRevB.97.201407} values to DFT}.

{The intervalley polaron mixes bright and momentum-forbidden excitonic states and does not affect the optical selection rules in no-phonon transitions, but polaron formation could influence the spin and valley relaxation.}

\textit{Conclusion.} 
We have developed the theory of a novel type of polaron corresponding to phonon-induced coupling between the two non equivalent valleys of transition-metal dichalcogenide monolayers. This inter-valley polaron associated to short-wavelength phonons results in both an increase of the electron effective mass and a decrease of the conduction band spin-orbit splitting. In the presence of an external magnetic field it will also lead to the renormalization of the cyclotron energy and the Landau level splitting. In contrast the intervalley polaron has negligible effects on the valence band due to its much larger spin-orbit splitting.

\textit{Acknowledgements.}
We are grateful to H. Dery for very stimulating discussions. M.A.S. and M.M.G. acknowledge partial support from LIA ILNACS through the RFBR project 17-52-16020. M.M.G. was partially supported by the RFBR project 17-02-00383. M.A.S. also acknowledges partial support of the Government of the Russian Federation (Project No. 14.W03.31.0011 at
the Ioffe Institute). We acknowledge funding from ANR 2D-vdW-Spin, ANR VallEx and ANR MagicValley. X.M. also acknowledges the Institut Universitaire de France.


\begin{thebibliography}{48}
\bibitem [{mas()}]{mass}%
  \BibitemOpen
  \href@noop {} { {\bibinfo {title} {We neglect the difference of the
  electron effective masses in the two spin subbands due to spin-orbit coupling
  and $\bm k\cdot \bm p$ mixing.}}}\BibitemShut {Stop}%
\bibitem [{shi()}]{shift}%
  \BibitemOpen
  \href@noop {} { {\bibinfo {title} {For the sake of simplicity, the
  overall shift of the conduction band $\propto \beta_{\bm K}
  \ln{(E_Q/\Delta_{cb})}$ is disregarded.}}}\BibitemShut {Stop}%
\end{thebibliography}
\end{document}

%\newcount\timehh  \newcount\timemm
%\timehh=\time \divide\timehh by 60
%\timemm=\time
%\count255=\timehh\multiply\count255 by -60 \advance\timemm by \count255

\title{Supplemental Information for\\
Intervalley Polaron in Atomically Thin Transition Metal Dichalcogenides}
\author{M.M. Glazov, M.A. Semina} 
\affiliation{Ioffe Institute, 194021 St. Petersburg, Russia}

\author{C. Robert, B. Urbaszek, T. Amand, X. Marie}
\affiliation{Universit\'e de Toulouse, INSA-CNRS-UPS, LPCNO, 135 Av. Rangueil, 31077 Toulouse, France}

%\\date{\today, \jobname.tex, printing time = \number\timehh\,:\,\ifnum\timemm<10 0\fi \number\timemm}

\maketitle

\tableofcontents

\section{Intravalley polaron}

In this section we study electron-optical phonon polaron in TMD monolayers. Here we consider only transitions with the intermediate states in the same valley. This introductory part is needed in order to introduce definitions and to justify the first-order calculation of the intervalley polaron effect.

In TMD monolayers the intravalley electron-phonon interaction can be provided by two mechanisms. First, there is the Fr\"ohlich interaction between the electron and dielectric polarization induced by the optical phonons in the monolayer~\cite{PhysRevB.85.115317,PhysRevB.94.085415,7496798} which in two-dimensional materials has a non-divergent form with the scattering matrix element 
\begin{equation}
\label{Fr}
M_{\bm q}^{(\rm F)} = \frac{C_{\rm F}}{1+2\pi \alpha q}, 
\end{equation}
where $\bm q$ is the phonon wavevector, $C_{\rm F}$ is bare Fr\"ohlich interaction constant, and $\alpha$ is the polarizability of the layer~\cite{PhysRevB.94.085415,7496798}. Also, the deformation potential interaction is possible with~\cite{PhysRevB.85.115317,PhysRevB.90.045422}
\begin{equation}
\label{Dp}
M_{\bm q}^{(\rm DP)} = D_{opt}.
\end{equation}
In this model the optical phonon deformation potential is taken to be independent of $\bm q$ (for wavevectors smaller than the Brillouin zone size). In both cases, the matrix element is regular at $\bm q\to 0$ and in many situations one can consider the coupling as $q$-independent.

\subsection{Perturbative derivation}

Here we derive the polaron shift and the correction to the effective mass of the electron for the intravalley, Fr\"ohlich-like, polaron in TMD ML. We consider dispersionless optical phonon with the frequency $\Omega_0$ and disregard the wavevector dependence of the matrix element replacing $M_q$ by the constant $M_0$.

It is convenient to measure energies in the units of optical phonon energy, $\hbar\Omega_0$, and present the electron momentum in the dimensionless form
\[
\bm w = \frac{\bm p}{\sqrt{2m\hbar\Omega_0}}.
\]
We follow the method of Ref.~\cite{Smondyrev1986} in order to analyze the polaron corrections in the second and fourth order perturbation theory. In the dimensionless units the second-order (with respect to $M_{\bm q}$) self-energy reads
\begin{equation}
\label{sigma:1}
\Sigma_1 = - \frac{\beta}{(2\pi)^2} \int  \frac{d\bm k}{1 - 2\bm w\cdot \bm k + k^2}.
\end{equation}
Here 
\begin{equation}
\beta = 2m\hbar\Omega_0 \mathcal S |M_0/\hbar\Omega_0|^2/\hbar^2,
\end{equation}
is the dimensionless coupling constant, $\bm k$ is the dimensionless phonon momentum. Integral in Eq.~\eqref{sigma:1} formally diverges (this is due to the fact that we neglect the momentum dependence of the electron-phonon matrix element). The electron energy renormalization in the first order in $\beta$ reads
\begin{equation}
\label{E1}
\mathcal E_1 = - \frac{\beta}{4\pi} \left( 2\ln{Q} + w^2\right),
\end{equation}
where $\bm Q$ is the cut-off wavevector. The correction to the effective mass is finite and the renormalized mass reads
\begin{equation}
\label{m1}
m^*_1 = m\left(1+ \frac{\beta}{4\pi}\right) = m\left(1+\frac{m\mathcal S |M_0|^2}{2\pi \hbar^3\Omega_0}\right).
\end{equation}
This result is in agreement with Eq.~(6) of the main text at $\Delta_{cb}=0$.

Equation~\eqref{E1} can be applied also to a valence band with corresponding parameters, see Ref.~\cite{Xiao_2017} where the effect for intravalley Fr\"ohlich polaron was discussed. The intravalley polaron corrections to the dispersion of the conduction and valence have the same order of magnitude~\cite{2019arXiv190102567V}.

For Fr\"ohlich polaron in three dimensions the higher order corrections in the coupling constant are small in the weak coupling regime. We demonstrate that this is also the case for the TMD MLs. In the fourth-order in electron-phonon interaction we have three contributions
\begin{widetext}
\begin{align}
&\Sigma_{2}^{(1)} = -\frac{\beta^2}{(2\pi)^4} \int  \frac{d\bm k d\bm q}{(1 - 2\bm w\cdot \bm k + k^2)^2(2 - 2\bm w\cdot \bm q + q^2)}, \\
&\Sigma_{2}^{(2)} = -\frac{\beta^2}{(2\pi)^4} \int  \frac{d\bm k d\bm q}{(1 - 2\bm w\cdot \bm k + k^2)[2 - 2\bm w\cdot (\bm k+\bm q) + (\bm k+\bm q)^2](1 - 2\bm w\cdot \bm q + q^2)}, \\
&\Sigma_{2}^{(3)} = \frac{\beta^2}{(2\pi)^4} \int  \frac{d\bm k d\bm q}{(1 - 2\bm w\cdot \bm k + k^2)^2(1 - 2\bm w\cdot \bm q + q^2)}, 
\end{align}
\end{widetext}
stemming from ``ladder'', ``crossing'' and ``disconnected'' diagrams. Although each of $\Sigma_{2}^{(1)}$ and $\Sigma_{2}^{(3)}$ diverge logarithmically, their sum is finite
\begin{equation}
\label{E2:13}
\mathcal E_2^{(1+3)} = \frac{\beta^2\ln{2}}{16\pi^2}  +\frac{\beta^2(\ln{2}+1/2)}{16\pi^2} w^2,
\end{equation}
The self-energy $\Sigma_2^{(2)}$ can be integrated numerically only (after analytical integration over the angles of $\bm k$ and $\bm q$ we arrive at the bulky expression which needs to be numerically integrated over the absolute values of the wavevectors):
\begin{equation}
\label{E2:13}
\mathcal E_2^{(2)} \approx -\frac{\beta^2}{16\pi^4} \left( 18.08 + 13.98 w^2\right),
\end{equation}
As a result, up to the second order we have
\begin{equation}
\label{E12}
\mathcal E_{1,2} = - \frac{\beta}{4\pi} \left( 2\ln{Q} + w^2\right) -\beta^2 \left(0.0072+0.0014w^2\right).
\end{equation}
The mass enhancement is finally given by
\begin{equation}
\label{m2}
m^* = m\left(1+ \frac{\beta}{4\pi}+0.0077\beta^2\right).
\end{equation}

\subsection{Feynman variational approach}

Following Refs.~\cite{PhysRev.97.660,MATSUURA19821471,HUYBRECHTS197895,PhysRevB.31.3420} we obtain the following expression for the variational functional (derived in the path-integral approach)
\begin{equation}
\label{Feynm:E}
E\{V,W\} = \frac{(V-W)^2}{2V} - \beta \int_0^\infty du e^{-u} \mathcal I(u;V,W),
\end{equation}
\[
\mathcal I(u;V,W) =  \frac{1}{2(2\pi)^2} \int d\bm k \exp{\left[-F(u;V,W) k^2  \right]}, 
\]
\[ 
F(u;V,W) = \frac{W^2}{2V^2}u + \frac{V^2-W^2}{2V^3} (1-e^{-uV}).
\]
where $V$ and $W$ are the variational parameters whose minimization $(V^*,W^*)$ yields the polaron shift. Equations above are valid for the polaron at rest. The renormalization of the effective mass can be readily calculated as
\begin{equation}
\label{Feynm:m}
\frac{m^*}{m} = 1 + \beta \int_0^\infty du e^{-u} \mathcal I'(u;V^*,W^*)
\end{equation}
 \[
\mathcal I'(u;V^*,W^*) =  \frac{1}{2(2\pi)^2} \int d\bm k \exp{\left[-F(u;V^*,W^*) k^2  \right]} \frac{k^2 u^2}{2}.
\]
In the linear in $\beta$ regime one can put $V^*=W^*$, thus $F=u/2$ and Eqs.~\eqref{Feynm:E}, \eqref{Feynm:m} pass to Eq.~\eqref{E1} and \eqref{m1}. In the second order in $\alpha$ we obtain (minimization yields $V^*=(1+\epsilon)W^*$, $\epsilon\ll 1$, $W^*\approx 7.577$)
\begin{equation}
\label{res:F}
\mathcal E = -\frac{\beta \ln{Q}}{2\pi}-0.0034\beta^2, \quad \frac{m^*}{m} = 1 + \frac{\beta}{4\pi} + 0.0069\beta^2.
\end{equation}
%At large $\beta$ one has $V^* \gg W^*$, as a result,
%\begin{equation}
%\label{approx:b:large}
%F = \frac{1}{2V}, \quad \mathcal I = \frac{\beta V}{4\pi}(1-e^{-Q^2/V}).
%\end{equation}

The analysis above demonstrates that the fourth-order (with respect to $M_{\bm q}$) corrections are indeed small. Therefore, in the remaining parts we will consider the second-order (with respect to $M_{\bm q}$, i.e., linear in $\beta$) corrections only.

\section{Intervalley polaron}

% For small $E_{\bm p}$ we can develop Eq.~\eqref{self:inter} into series in $E_p$ with the same result as Eq.~\eqref{disper:inter:1}.

\subsection{Renormalization of Landau levels}

The correction to the energy of electron on $n$-th Landau level is given, in the second-order in the intervalley electron-phonon interaction ($\propto |M_{\bm K}|^2$ or $\propto \beta_{\bm K}$), by
\begin{equation}
\label{Landau}
\delta E_{n,\pm} = \sum_{\bm q, n'k_y'} \frac{|M_{\bm K}|^2 \left|\langle n'k_y'|e^{\mathrm i \bm q \bm r}|n k_y\rangle\right|^2}{\hbar\omega_c (n-n') - \hbar\Omega_{\bm K} \pm \Delta_{cb} },
\end{equation}
where $\omega_c = |eB_z/mc|$ is the cyclotron frequency, $|n k_y\rangle$ is the Landau level wavefunction, and $\bm q$ is the phonon wavevector (minus $\bm K$, the intervalley wavevector).  We use the Landau gauge with the vector potential $\bm A = [0, B_z x,0]$; $z$ is the monolayer normal, $x$ and $y$ are the axes in the sample plane. As discussed in the main text, we neglect the phonon dispersion and the dependence of the electron-phonon interaction matrix element on the phonon wavevector. The matrix element of the transition between the Landau levels can be expressed as~\cite{gurevich:firsov}
\begin{equation}
\label{me:Landau}
\langle n'k_y'|e^{\mathrm i \bm q \bm r}|n k_y\rangle = \delta_{k_y',k_y+q_y} Q_{n}^{n'-n}(u),
\end{equation}
where $u = q^2l_B^2/2$ and $l_B = \sqrt{\hbar c/|eB_z|}$ is the magnetic length,
\[
Q_{n}^{n'-n}(u) = \frac{e^{-u/2} u^{(n'-n)/2}}{\sqrt{n'!n!}} L_n^{(n'-n)}(u), 
\]
and 
\[
L_n^{m}(u) = e^u u^{-m}  \frac{d^n}{du^n} (e^u u^{n+m}), 
\]
is the generalized Laguerre polynomial. Here we assumed $n'>n$, similar expressions can be written for $n'\leqslant n$. 

The summation over $k_y'$ is removed by using the $y$-component of momentum conservation $\delta$-function and the summation over $\bm q$ can be performed by virtue of 
\[
\int_0^\infty du \left|Q_{n}^{n'-n}(u) \right|^2 = 1,
\]
yielding
\begin{equation}
\label{Landau:1}
\delta E_{n,\pm} = \frac{\mathcal S |M_{\bm K}|^2}{2\pi l_B^2} \sum_{n'=0}^\infty \frac{ 1}{\hbar\omega_c (n-n') - \hbar\Omega_{\bm K} \pm \Delta_{cb} }.
\end{equation}
As above, this sum is divergent logarithmically, but we are interested only in the renormalization of Landau level distances. Thus, the correction to $\hbar\omega_c$ for the transition between $n$-th and $n-1$-th Landau levels reads
\begin{widetext}
\begin{equation}
\label{Landau:2}
\delta E_{n,\pm} - \delta E_{n-1,\pm} = -\frac{\beta_{\bm K} \hbar\omega_c}{4\pi} \frac{\hbar\Omega_{\bm K}+ \Delta_{cb}}{\hbar\Omega_{\bm K}\mp \Delta_{cb}} \sum_{n'=0}^\infty \left[\frac{ 1}{1+\eta_\pm(n'-n)} -\frac{ 1}{1+\eta_\pm(n'-n+1)}  \right] = -\frac{\beta_{\bm K} \hbar\omega_c}{4\pi} \frac{\hbar\Omega_{\bm K}+ \Delta_{cb}}{\hbar\Omega_{\bm K}\mp \Delta_{cb}} \frac{1}{1-\eta_\pm n},
\end{equation}
\end{widetext}
where $\eta_\pm = \hbar \omega_c/(\mp \Delta_{cb} + \hbar\Omega_{\bm K})$. Note that in order to calculate the sum one has to make a replacement $n'\to n+1$ in the second term. Equation~\eqref{Landau:2} is in agreement with Eq.~(9) of the main text.

\subsection{Relation to Holstein polaron}

It is instructive to discuss the difference between our model and the Holstein polaron~\cite{HOLSTEIN1959325,Alexandrov2010} where the polaron radius is on the order of lattice constant. Indeed some measurements of strong band gap renormalization in surface doped thick layers of MoS$_2$ were interpreted recently on the basis of Holstein polaron~\cite{kang2018holstein}. Although in our case, the involved wavevectors of phonons correspond to the BZ size and the phonon wavelengths are on the order of the lattice constant, the direct analogy with the Holstein polaron is not so relevant: In the Holstein polaron approach the polaron energy $E_{pol}$ roughly corresponds to the band width. In contrast, in our case the weak coupling regime is realized: $\beta_{\bm K} \lesssim 1$ thus, the polaron energy $E_{pol}\sim \beta_{\bm K} \hbar\Omega_{\bm K}$ is negligible compared with the band width. The calculation of intervalley Holstein polaron characteristics in ML TMDs is beyond the scope of this paper.

\section{Symmetry of the involved phonon modes}

\begin{table}[h]
\caption{Irreducible representations of the $C_{3h}$ point group and characters of the relevant elements, $\omega = \exp{(2\pi \mathrm i /3)}$. The center of the point group transformations corresponds to the center of hexagon.}\label{tab:irreps:C3h}
\begin{ruledtabular} 
\begin{tabular}{c|c|c|c|c|c|c}
 & & $e$ & $C_3$ & $S_3$ &  $\sigma_h$ & Basic functions\\
 \hline\hline
 $A'$ & $K_1$ & $1$ & $1$ & $1$ & $1$ & $x^2+y^2$; $z^2$\\
 \hline
 $E_1'$ & $K_2$ & $1$ & $\omega$ & $\omega$ & $1$ & $x+\mathrm i y$; $(x-\mathrm i y)^2$\\
  \hline
 $E_2'$ & $K_3$ & $1$ & $\omega^2$ & $\omega^2$ & $1$ & $x-\mathrm i y$; $(x+\mathrm i y)^2$\\  
 \hline
  $A''$ & $K_4$ & $1$ & $1$ & $-1$ &$-1$ &  $z$\\
\hline
 $E_1''$ & $K_5$ & $1$ & $\omega$ & $-\omega$ & $-1$ & $S_x+\mathrm i S_y$; $(S_x-\mathrm i S_y)^2$\\
  \hline
 $E_2''$ & $K_6$ & $1$ & $\omega^2$ & $-\omega^2$ & $-1$ & $S_x-\mathrm i S_y$; $(S_x+\mathrm i S_y)^2$\\  
\end{tabular}
\end{ruledtabular}
\end{table}

We consider phonons which provide the electron mixing between the $\bm K$ and $\bm K'$ points of the BZ. Note that in the $\bm K$ valley the interband optical transition is allowed in $\sigma^+$ polarization and in the $\bm K'$ valley is in the $\sigma^-$ polarization. The wavevector of such phonon equals to $\bm K$, see inset in Fig.~1 of the main text. Note that for simplicity we consider the phonon absorption process and the transfer $\bm K\to \bm K'$, the analysis of the emission process is similar, but requires to consider the phonons in the $\bm K'$ point of the BZ. In order to determine the selection rules it is needed to consider the corresponding wavevector group $C_{3h}$. Since we neglect the spin mixing, it is sufficient to consider only transformation of the orbital parts of the Bloch functions of electrons. Thus, disregarding the spin and spin-orbit coupling the basic functions of the conduction band top transform according to $\mathcal D_{\bm K} =K_2$ ($E_1'$) in the $\bm K$ valley and according to $\mathcal D_{\bm K'} = K_3$ ($E_2'$) in the $\bm K'$ valley.

It follows from Ref.~\cite{PhysRevB.90.115438} that the phonons at the $\bm K$ point of the BZ transform according to the following irreducible representations of the $C_{3h}$ point group ($2Hc$-polytype, $N=1$):
\begin{equation}
\label{phonons:irreps}
2K_1,~K_2,~2K_3,~K_4,~2K_5,~K_6.
\end{equation}
There are $9$ phonon modes in total ($3$ atoms in the unit cell). Note that in the $\bm K'$ point the phonon modes are complex conjugate, $2K_1,~2K_2,~K_3,~K_4,~K_5,~2K_6$.
In order to enable the electron mixing, the symmetry of the phonon mode $\mathcal D_{ph}$ should be such that the direct product of the representations
\begin{equation}
\label{prod:irreps}
\mathcal D_{\bm K'}^* \otimes \mathcal D_{ph} \otimes \mathcal D_{\bm K} = K_3\otimes \mathcal D_{ph},
\end{equation}
contains the identical one, $K_1$ (or $A'$). This is possible only for the phonon of $K_2$ symmetry. Thus, only one phonon mode is allowed for such a transfer~\cite{PhysRevB.85.115317,Song:2013uq,Carvalho:2017aa,PhysRevLett.115.115502,Zhu579}.

%merlin.mbs apsrev4-1.bst 2010-07-25 4.21a (PWD, AO, DPC) hacked
%Control: key (0)
%Control: author (0) dotless jnrlst
%Control: editor formatted (1) identically to author
%Control: production of article title (0) allowed
%Control: page (1) range
%Control: year (0) verbatim
%Control: production of eprint (0) enabled
%

%\bibliography{polaron}